\documentclass[english,twocolumn, superscriptaddress, amsmath]{revtex4}
\usepackage[colorlinks=true,linkcolor=blue,citecolor=blue]{hyperref}
\usepackage{times}
\usepackage{amsmath}
\usepackage{amssymb}
\usepackage{amsthm}
\usepackage{amsfonts}
\usepackage{enumerate}
\usepackage{latexsym}
\usepackage{ifpdf}
\usepackage{soul}
\usepackage{graphicx}
\usepackage{makeidx}
\usepackage{color}
\newcommand{\SRO}{SrRuO$_3$ }
\newcommand{\CRO}{CaRuO$_3$ }
\newcommand{\STO}{SrTiO$_3$ }

\usepackage{graphicx}% Include figure files
\usepackage{dcolumn}% Align table columns on decimal point
\usepackage{bm}% bold math

\usepackage{babel}

\begin{document}

\title{Effect of microstructure on the electronic transport properties of epitaxial \CRO thin films}
\author{Gopi Nath Daptary}

%\affiliation{Department of Physics, Indian Institute of Science, Bangalore 560012, India}
\author{Chanchal Sow}
\author{Suman Sarkar}
%\affiliation{Department of Physics, Indian Institute of Science, Bangalore 560012, India}

\author{Santosh Chiniwar}
%\affiliation{Center For Nano Science And Engineering, Indian Institute of Science, Bangalore 560012, India}
\author{P. S. Anil Kumar}
%\affiliation{Department of Physics, Indian Institute of Science, Bangalore 560012, India}
\address{Department of Physics, Indian Institute of Science, Bangalore 560012, India}

\author{Anomitra Sil}
\address{Center For Nano Science And Engineering, Indian Institute of Science, Bangalore 560012, India}
\author{Aveek Bid}
\email{aveek@iisc.ac.in}
\affiliation{Department of Physics, Indian Institute of Science, Bangalore 560012, India}

\begin{abstract}
	We have carried out extensive comparative studies of the structural and transport properties of \CRO thin films  grown under various oxygen pressure. We find that the preferred orientation and surface roughness of the films are strongly affected by the oxygen partial pressure during growth. This in turn affects the electrical  and magnetic properties of the films. Films grown under high oxygen pressure have the  least surface roughness and show transport characteristics of a good metal down to the lowest temperature measured. On the other hand, films grown under low oxygen pressures    have high degree of surface roughness and show signatures of ferromagnetism. We could verify that the low frequency resistance fluctuations (noise) in these films arise due to thermally activated fluctuations of local defects and that the defect density matches with the level of disorder seen in the films through structural characterizations.
\end{abstract}

%\begin{keyword}
%A. Thin films; B. Strain; C. Resistivity; D. Magnetoresistance; E. Hysteresis F. Resistance fluctuation
%\end{keyword}

%\end{frontmatter}
\maketitle

\section{Introduction}

Complex perovskite oxides of the $4d$ transition ruthenates have attracted interest in condensed matter physics due to the strong correlation between their electronic structure, magnetic structure and transport properties~\cite{mazin1997electronic,santi1997calculation,park2004electronic,cao1997thermal}. 
Of special interests are the members of the Ruddlesden-Popper series ~\cite{ruddlesden1957new} of ruthenates $A_{n+1}Ru_{n}O_{3n+1}$ where $n$ denotes the number of Ru-O layers between two alternative layers of A-O. The interest in these materials stems form the fact that their band structure is very strongly influenced by lattice distortions. A classic example is the case of \SRO and \CRO which are the infinite dimensional members (n =$\infty$) of the Ruddlesden-Popper series. At low temperatures \SRO is a ferromagnetic metal. Iso-structural and iso-electronic \CRO on the other hand is at the border between metal and non-metal. Until recently, it was thought that \CRO had no magnetic order. Recent experiments have shown that this is not necessarily true. Depending on its growth parameters \CRO thin films can show signatures of either long range magnetic order like that in \SRO or short range interactions, possibly a spin-glass like state~\cite{kolev2002raman,felner2000caruo}. The transport properties of \CRO thin films are equally intriguing. Conventional Fermi-liquid behaviour does not hold well in \CRO~\cite{klein1999possible,cao2008non}. The temperature range over which this non-Fermi liquid like behaviour sets in and the exact values of the exponents describing this behaviour are still under debate. A survey of the existing data on \CRO clearly shows that like most oxide films, the transport properties depend very strongly on the quality of the films {~\cite{tripathi2014ferromagnetic,cao2008non,lee2002non} and till date there is no clear study of the effect of growth parameters on the structural, magnetic and transport properties of this material. Motivated by this observation we have studied the correlation between structural and electrical transport properties of \CRO thin films grown under different deposition conditions.
	
	\section{ Methods}
	
	Epitaxial thin films of \CRO (001) (CRO) were grown on single crystal \STO(001) (STO) substrates by Pulsed Laser Deposition (PLD) technique using a KrF ($\lambda=248$ nm) laser.  A stoichiometric target (20 mm diameter) of \CRO prepared using solid state reaction process was used as a target for the PLD. The as procured  \STO substrates were cleaned in trichloroethylene, acetone, isopropyl alcohol and finally nitrogen gas spray. The cleaned substrates were annealed in-situ at $700^{\circ}{\rm C}$ for 1 hour in oxygen atmosphere before the deposition. The deposition conditions were:  (i) 1.5 J/$cm^2$ fluence, (ii) frequency=5 Hz and (iii) $700^{\circ}{\rm C}$ substrate temperature. The only parameter varied during the growth was the oxygen partial pressure. Multiple films in each group were studied - in this letter we report the results  on three representative films, : S1 (0.15 mbar oxygen pressure), S2 (0.30 mbar oxygen pressure) and S3 (0.45 mbar oxygen pressure). The deposition rate in each case was 1.5 nm/min. Post-deposition, all the films were in-situ annealed at $500^{\circ}{\rm C}$ in  500 mbar oxygen pressure for 30 minutes. This is a commonly used method in the growth of many oxide thin films to improve their oxygen content~ \cite{cancellieri2010influence,biscaras2012irreversibility,beshkova2009effect}. Generally, in oxides thermogravimetric analysis shows that there is an oxygen intake between $450^{\circ}{\rm C}$ and $500^{\circ}{\rm C}$. Hence the samples were annealed at  $500^{\circ}{\rm C}$ to have the minimum oxygen deficiency. The thicknesses of the films as measured were 30$\pm$2~nm.
	
	%%------------------------------
	\begin{figure}[t]
		\begin{center}
			\includegraphics[width=0.48\textwidth]{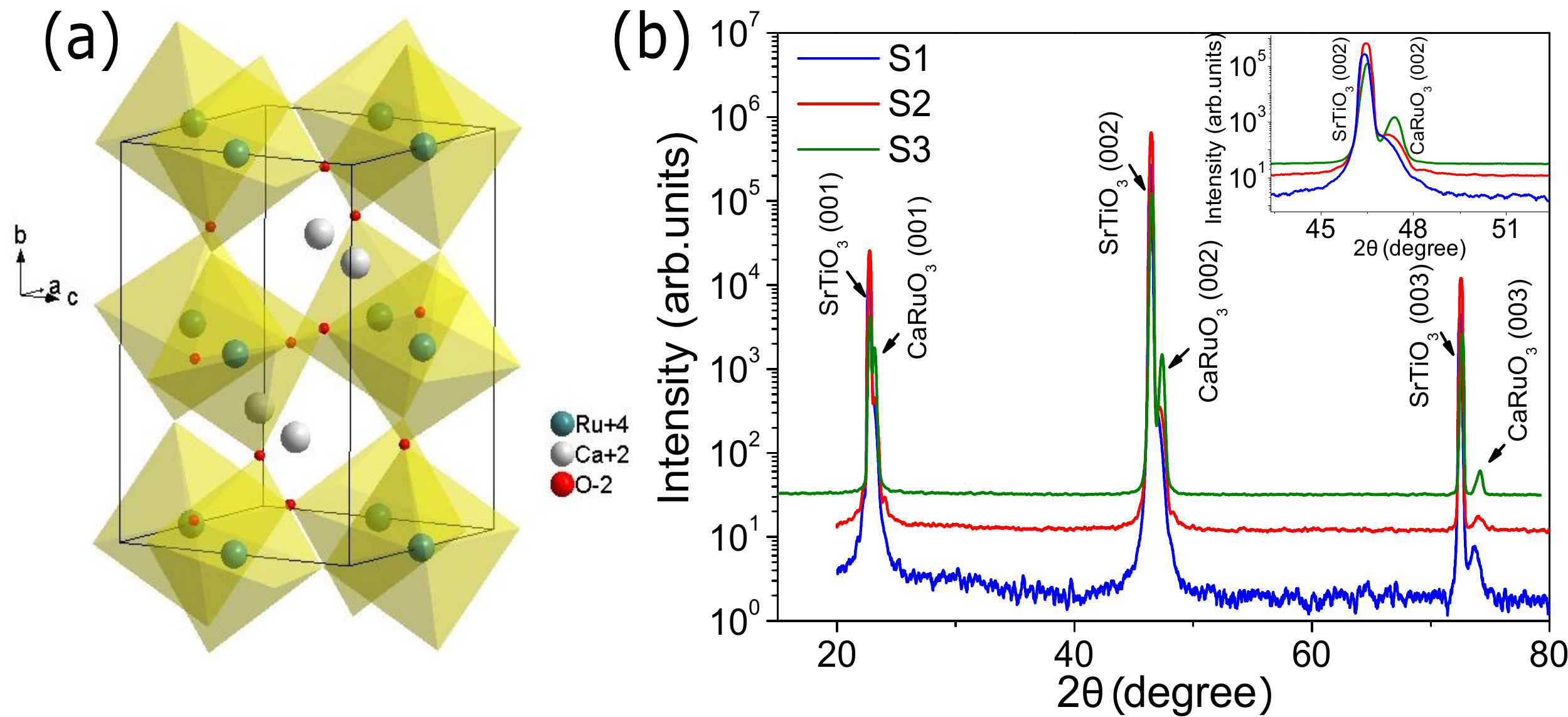}
			\small{\caption{(a) The atomic structure of \CRO film. (b) Plot of $\theta$-2$\theta$ x-ray diffraction pattern of the three \CRO films. The inset shows the (002) x-ray diffraction peaks - the data have been shifted vertically for clarity.\label{fig:xrd}}}
		\end{center}
	\end{figure}
	%%_--------------------------------

	\section{Results}

	The structural properties of the three films were characterized using high resolution x-ray diffraction (HRXRD), Scanning electron microscope (SEM) and Atomic force microscope (AFM) measurements. In figure~\ref{fig:xrd}(a), we show the atomic structure of \CRO film and in figure~\ref{fig:xrd}(b), we show the $2\theta$ x-ray diffraction pattern (XRD) for the three representative films. The measurements were done at room temperature in a Rigaku SmartLab using a source wavelength of 1.5418\AA. In all three cases the XRD peaks of the films and that of the substrate match well with published data~\cite{higashi2001crystal}. From the measured peak positions, the lattice constants were evaluated and have been tabulated in table~\ref{tab:films}. To obtain an understanding of the nature of strain in these films Reciprocal Space Mapping (RSM) were performed for the (012) peaks  by high-resolution asymmetrical x-ray diffraction measurements on all three films - the results are plotted in figure~\ref{fig:RSM}. The dotted line in each graph denotes the direction of elongation of the reciprocal lattice point (RLP). To obtain the strain state of the films we have calculated the quantity $(S-F)/S$; where $S$ is the substrate lattice parameter and $F$ is the film lattice parameter. The amount of in-plane as well as out of plane strain of \CRO films could be quantified from the x-ray diffraction studies and RSM data - the results are tabulated in table~\ref{tab:films}.  From the data it is clear that the film S1 has the maximum in-plane strain while it is the least for film S3. This is in accordance with previous results where it was shown that low oxygen partial pressure results in the growth of ruthenium perovskites films with large lattice mismatch with SrTiO$_3$ substrate which in turn leads to high strain levels in the film.~\cite{RevModPhys.84.253}.

	\begin{figure*}[tbh]
		\begin{center}
			\includegraphics[width=\textwidth]{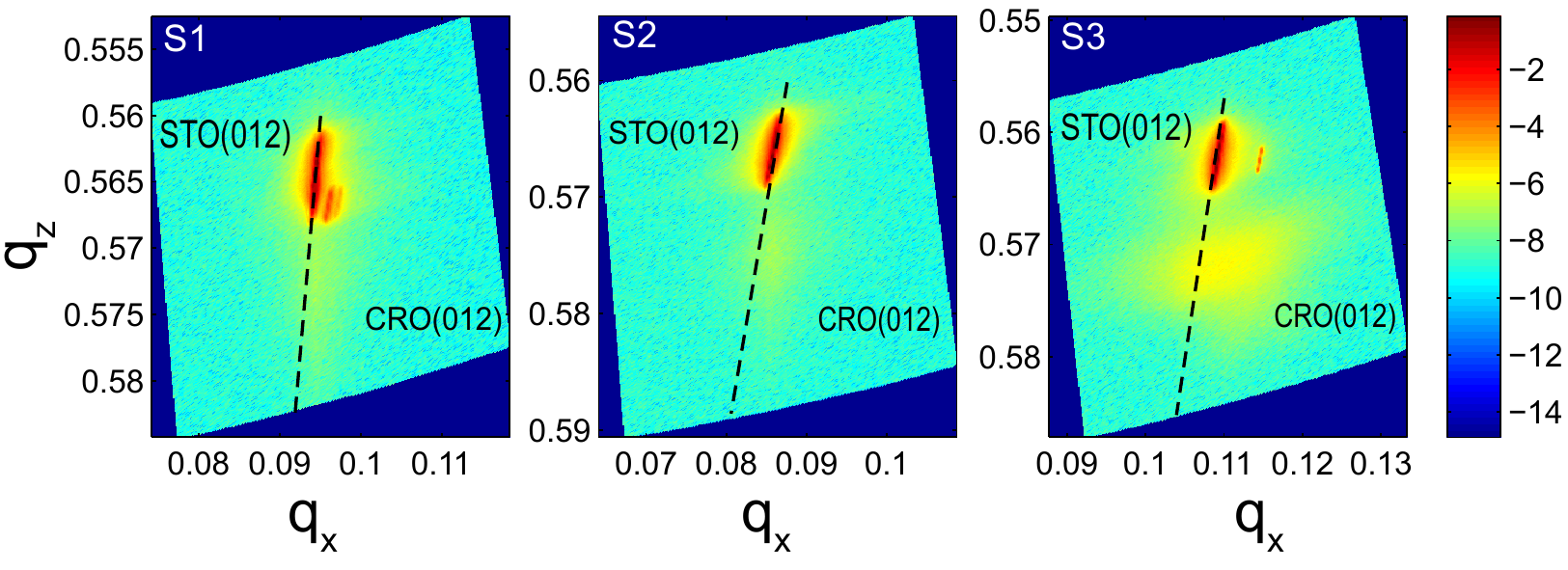}
			\small{\caption{ Plots of the RSM data of three films S1, S2 and S3. The intensities in all cases have been scaled by the maximum substrate intensity of S2. The dotted line in each graph denotes the direction of elongation of the resiprocal lattice point (RLP). } \label{fig:RSM}}
		\end{center}
	\end{figure*}

	Both scanning electron microscopy (figure~\ref{fig:SEM}(a-c)) Atomic force microscope (AFM) scans (figure~\ref{fig:SEM}(d-f)) show that the film S3 has the least surface roughness. The line-scans shown by the red line in each AFM image were used to calculate the rms surface roughness. The rms roughness of three films S1, S2 and S3 were 2.73$\pm$0.15, 1.81$\pm$0.12 and 0.817$\pm$0.07  nm respectively showing again that the film S3 was of the highest quality.
	
	From the analysis ofFull width at half maximum (FWHM) of the XRD peak and c/b ratio with oxygen pressures (see figure ~\ref{fig:FWHM}(a) and figure ~\ref{fig:FWHM}(b) ), it is clear that the quality of the films improves significantly with increasing oxygen partial pressure. It was also observed by us from AFM studies that the surface roughness of the films decreases with increasing oxygen pressure: this might suggest a change from island-like growth to layer-by-layer growth. A possible explanation for this might be that oxygen-pressure results in a decrease in the mean-free-path of species thus allowing to form a layer-by-layer growth. 
	
	We propose that there are no critical structural changes of the films. Crystallographic lattice parameters and X, Y, Z positions (atomic positions) were used to generate the stereographic projection for CaRuO3. Rocking curve measurements were performed on different off specular planes to confirm the crystal structure of the films It was observed that the generated stereographic projection has a good match with the crystal structure for all the three films. The slight shifts observed in the peak positions appeared because of strain. From these, we concluded that the strain in the films did not lead to significant structural changes.

	\begin{table*}[t]
		%\begin{ruledtabular}
		\centering
		
		\begin{tabular}{| l | c | c | c | c | c | c | c | c | }
			\hline
			film &b&c&in-plane strain&out-of-plane strain& $\rho$ &  $l_{mfp}$ & $k_Fl_{mfp}$ &$l_\phi$\\ \hline
			& \AA& \AA& \%&\% &$\mu\Omega-m$&$m$&   &$nm$                                               \\ \hline
			S1 &3.69 &3.9 &5.6 &0.26 &2.20 &0.8$\times10^{-9}$&6.595&49\\ \hline
			S2 &3.83 &3.86 &2.04 &1.28 &1.27 &1.4$\times10^{-9}$&11.40&91\\ \hline
			S3 &3.91 &3.83 &0 &2.04 &0.25 &7.0$\times10^{-9}$&58.42&-\\ \hline
			
		\end{tabular}
		\small{\caption{Values of the in-plane lattice parameter b, out-of-plane lattice parameter c, in-plane strain, out of plane strain, electrical resistivity $\rho$, charge carrier mean free path $l_{mfp}$ and the Ioffe-Regal parameter $k_Fl_{mfp}$ for the three films at 300 K. Included also are the values of the phase coherence length $l_\phi$ for the films S1 and S2 measured at 245~mK.\label{tab:films}}}
		%\end{ruledtabular}
		
	\end{table*}

	\begin{figure*}[tbh]
		\begin{center}
			\includegraphics[width=\textwidth]{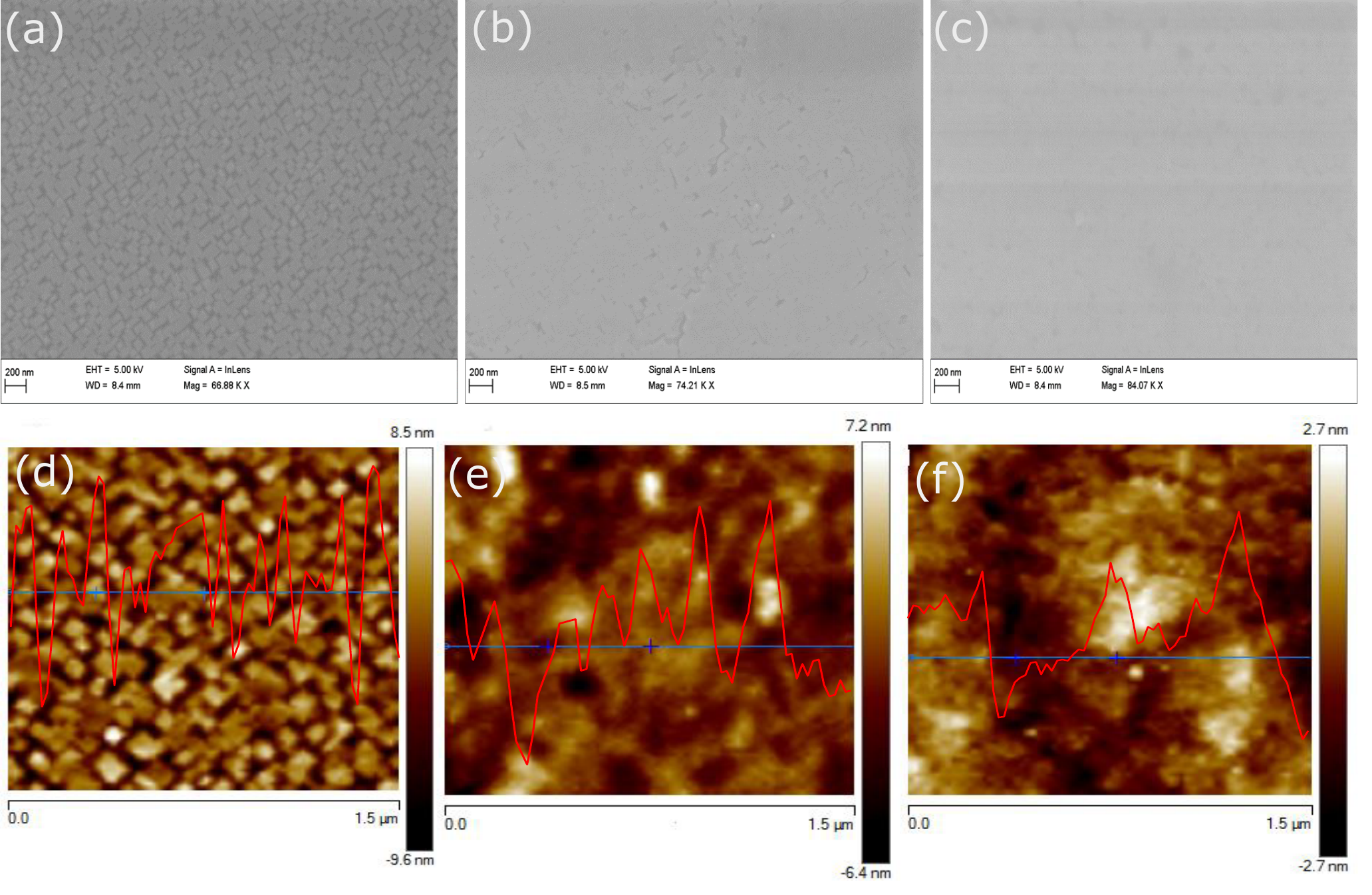}
			
			\small{\caption{ Top panel: Scanning Electron Microscope images of (a) film S1, (b) film S2 and (c) film S3 taken in a ZEISS-ULTRA 55. Lower panel: Atomic force microscope images of (d) film S1, (e) film S2 and (f) film S3 taken in a Bruker Dimension ICON AFM. The red lines in each AFM image is a typical line-scan from which the rms surface roughness of the three films were extracted. \label{fig:SEM}}}
		\end{center}
	\end{figure*}

	\begin{figure}[tbh]
		\begin{center}
			\includegraphics[width=0.48\textwidth]{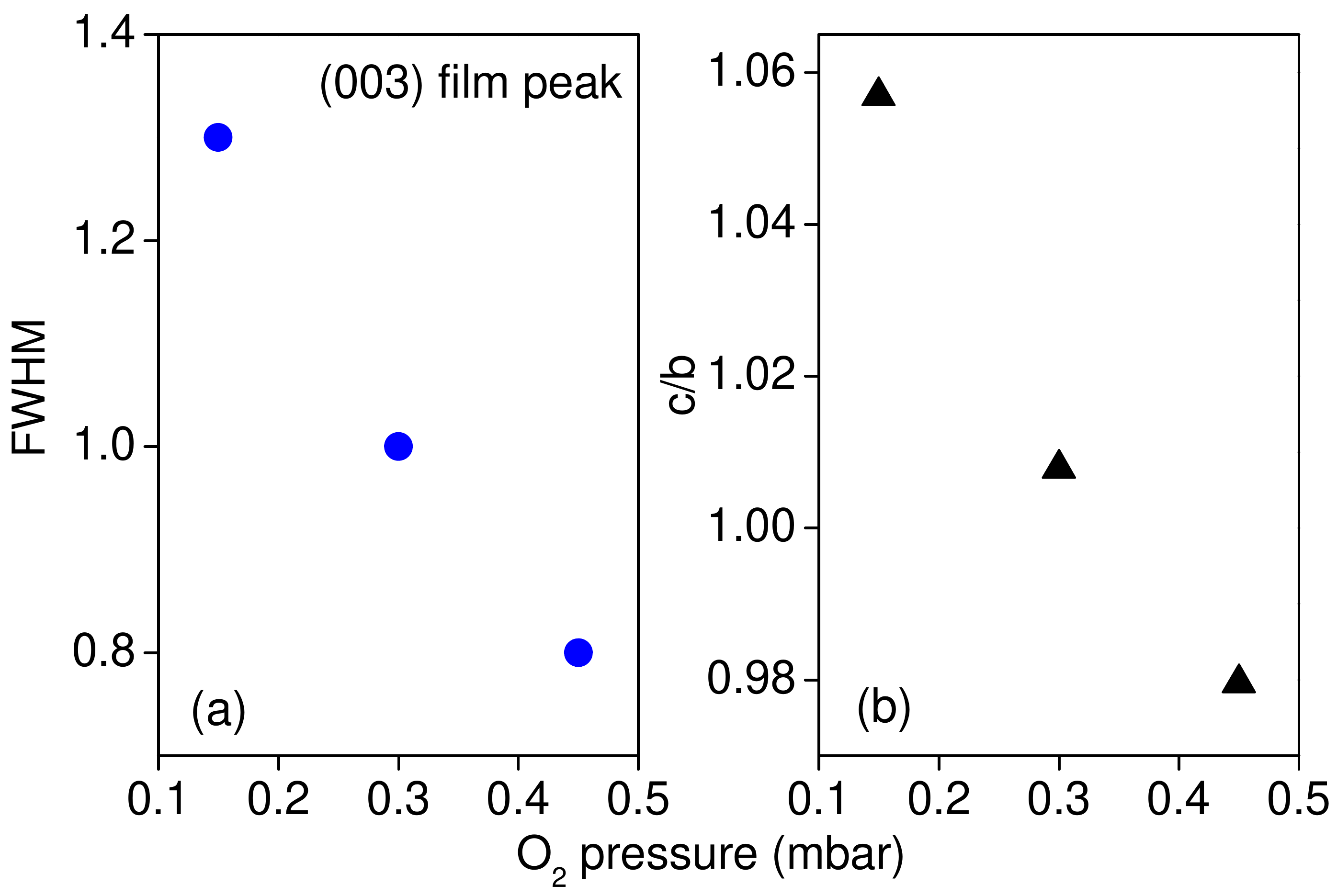}
			\small{\caption{ Plot of the (a) Full width at half maximum of the (003) peak and (b) $c/b$ ratio for the three films S1, S2 and S3 versus oxygen pressure.} \label{fig:FWHM}}
		\end{center}
	\end{figure}

	\begin{figure}[tbh]
		\begin{center}
			\includegraphics[width=0.48\textwidth]{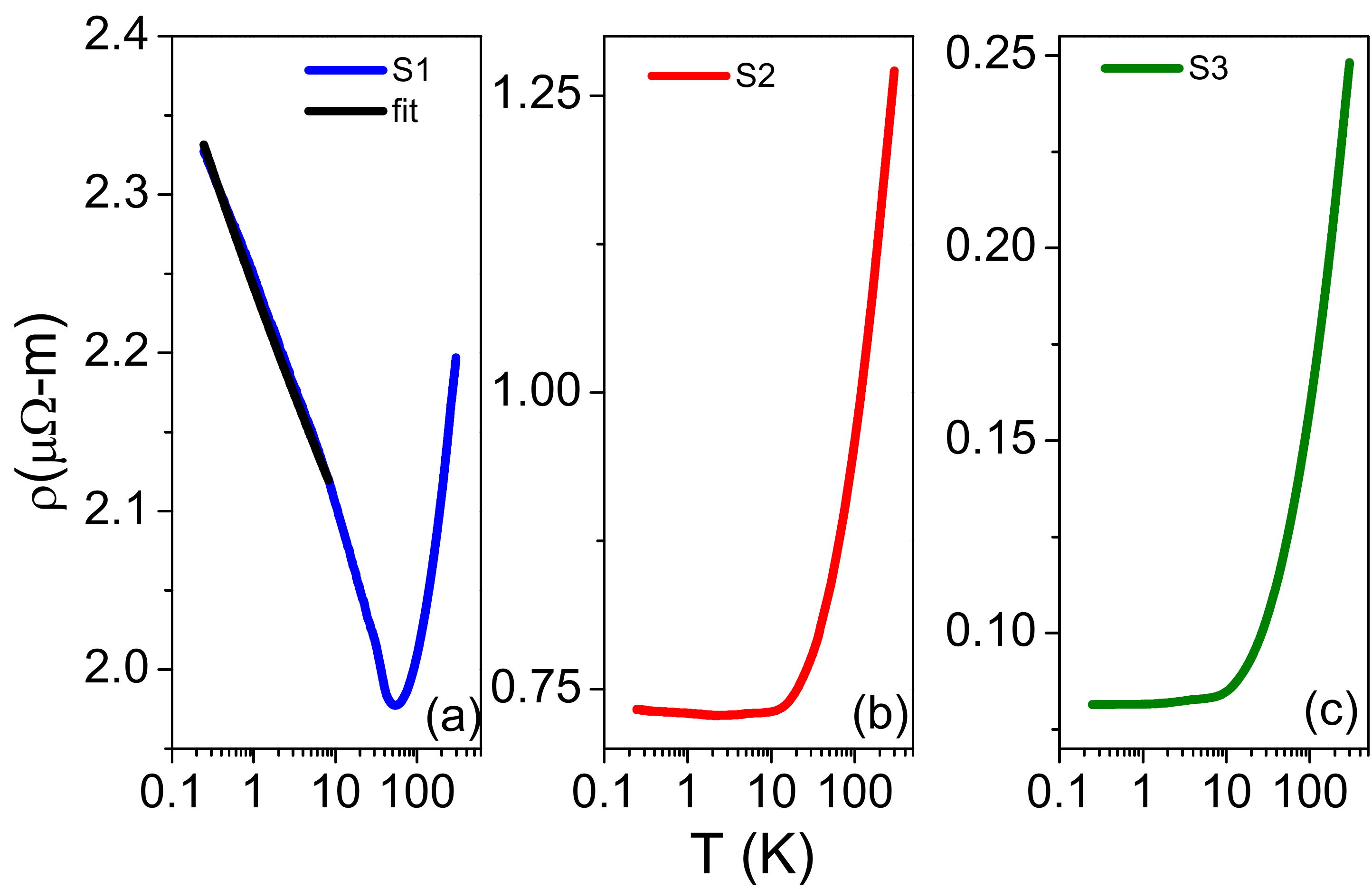}
			\small{\caption{ Resistivity of the three \CRO films as a function of temperature. The black solid line in (a) is the fit to the eqn.~\ref{WL}.\label{fig:resistivity}}}
		\end{center}
	\end{figure}
	%%-----------------------
	
	So far we have established from XRD, SEM and AFM studies that the film S3 grown under the highest oxygen partial pressure has the best structural properties in terms of lattice structure, surface voids density and surface roughness. In order to correlate these structural characteristics with the transport properties of the films we studied the resistance, magnetoresistance and resistance fluctuations of these films over an extensive temperature range from 295~K down to 245~mK. The resistance of the films were measured using a standard 4-probe ac lock-in technique. Linear electrical probes were defined by metaliziation (10~nm Cr/100~nm Au) through a metal mask. The measurements were performed in  a cryogenic He-3 system
	
	The resistivity as a function of temperature for the three different films is shown in figure~\ref{fig:resistivity}. Film S3 shows metallic behaviour (parametrized by $dR/dT>0$) down to 0.245~K. In contrast S1 shows a metallic behaviour till 50~K before its resistivity begins to logarithmically increase with further decrease in temperature. For sample~S2 the metallic behaviour persists down to about 2~K before it shows a small upturn in resistivity - the magnitude of the low temperature rise in resistivity for S2 being much smaller than that in S1. As expected, film S3 has the smallest room temperature resistivity and the largest residual resistivity ratio (RRR) as compared to S1 and S2 - it may be noted that the value of the room temperature resistivity for the films S3 is among the lowest reported  till date~\cite{tripathi2014ferromagnetic,cao2008non} attesting to the high quality of the film.  We have calculated the  Ioffe-Regal parameter $k_Fl_{mfp}$ to quantify the disorder level of three films by considering free electron approximation. The value of $k_F$ and $l_{mfp}$ are obtained from $k_F^3=3 \pi^2n$ and $l_{mfp}=v_F \tau$,  where $v_F$ and $\tau$ are Fermi wave vector and mean time respectively~\cite{ashcroft} and $n$ is the number density taken from ref.~\cite{tripathi2014ferromagnetic}.  The resistivity, charge carrier mean free path $l_{mfp}$ and the  Ioffe-Regal parameter $k_Fl_{mfp}$ for the three films at 300~K are tabulated in table~\ref{tab:films}. The very low value of $k_Fl_{mfp}$ measured in film S1 indicates the defective nature of the film and corroborates the conclusions reached from the structural studies.

	The low temperature upturn observed in the resistivity of a metallic film can have two possible origins - (1) weak localization (WL) or (2) electron-electron interactions (EEI). When spin-orbit scattering and spin-spin scattering are negligible, both WL and EEI give a logarithmic divergence to the resistance at low temperatures: 
	\begin{eqnarray}
	\frac{R(T)-R(T_0)}{R(T)R(T_0)} \propto ln\Bigg(\frac{T}{T_0}\Bigg)
	\label{WL}
	\end{eqnarray}
	The functional forms of the contributions from WL and EEI to the temperature dependence of resistance at zero-magnetic field being very similar, it is not possible to unambiguously establish from the resistivity data alone which of these processes dominate the electrical transport at low temperatures. However, it has been shown that the correction to the magnetoresistance at low magnetic fields from these two effects can be drastically different~\cite{hikami1980spin,jetp, Bergmann19841} and hence low field magnetoresistance (MR) measurements can be used to distinguish between the two mechanisms.
	
	%%_--------------------
	\begin{figure}[t!]
		\begin{center}
			\includegraphics[width=0.48\textwidth]{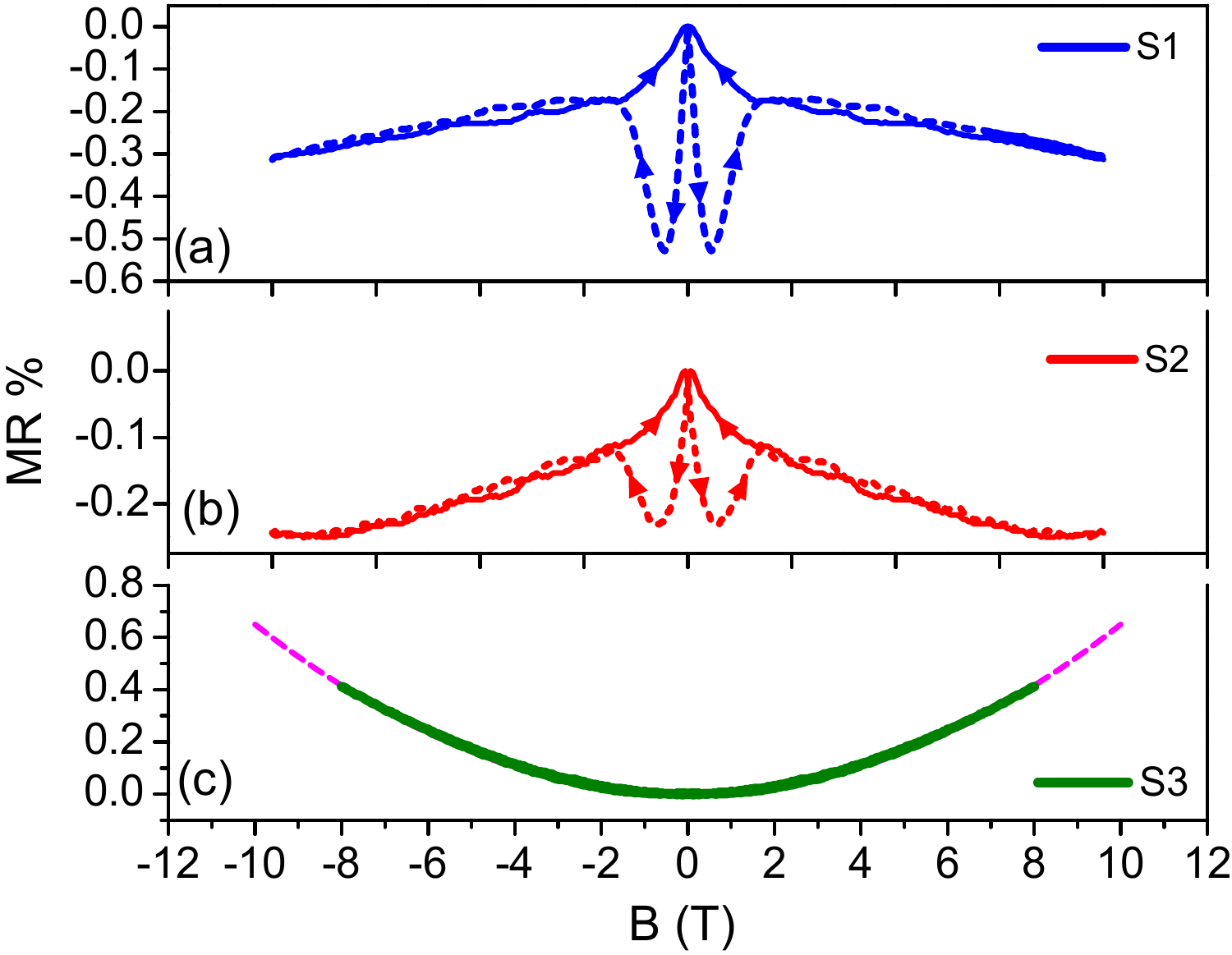}
			\small{\caption{ (a) and (b) show respectively the magnetoresistance (MR) of the films S1 and S2 measured at 245~mK. The dotted lines are the data for increasing magnetic field while the solid lines are the data for decreasing magnetic field.  (c) Solid olive line shows the magnetoresistance (MR) of the film S3 measured at 245~mK. The data for increasing and decreasing magnetic field superpose on each other.   The dotted line is a quadratic fit to the data.  \label{fig:mr}}}
		\end{center}
	\end{figure} 
	%%-----------------------------
	
	MR measurements were performed on all the three films at 245~mK in magnetic fields upto 8~T - the results are plotted in figure~\ref{fig:mr}. For both S1 and S2 the MR are negative and hysteretic with a butterfly pattern characteristic of MR seen in ferromagnetic materials ~\cite{mehta2012evidence,barzola2012revealing,bern2013structural} with the area of the hysteresis loop much larger for S1 as compared to that in S2. This is consistent with the previous observations of ferromagnetism in  tensile strained 30~nm films of CaRuO$_3$ grown on SrTiO$_3$ substrates~\cite{tripathi2014ferromagnetic}. The low field ($|B|<0.35$~T) magneto-conductance $\sigma(B)$ data for these two films  were fitted to the Hikami-Larkin-Nagaoka (HLN) equation~\cite{hikami1980spin}: 
	\begin{eqnarray}
	\sigma(B)-\sigma(0) = \alpha\frac{e^2}{2\pi^2\hbar} \Bigg(ln\Bigg(\frac{B_{\phi}}{B}-\psi(\frac{1}{2}+\frac{B_{\phi}}{B})\Bigg)\Bigg)
	\label{nln}
	\end{eqnarray}
	\noindent where $B_{\phi}$ is related to the phase coherence length $l_{\phi}$ as $B_{\phi}=\hbar/(4el_{\phi}^2)$. From the measured value of $l_{\phi}$ we verify that for both the films  $l_\phi > t$, where $t$ is the thickness of the film (see table~\ref{tab:films}). This justifies the use of the two-dimensional form of the equations for WL or EEI. It may be noted that EEI can also lead to MR, but only in much higher fields than are important in eqn.~\ref{nln}~\cite{Bergmann19841}, we can therefore ignore this contribution to the low-field MR.  The MR of the film S3, on the other hand has the classic quadratic behavior over the entire magnetic field range, characteristic of a good non-magnetic metal. 
	
	The emergence of signatures of FM order in \CRO films having a high degree of tensile strain can be understood using the following reasoning: the magnetic order of CaRuO$_3$ differs from its isoelectronic dual SrRuO$_3$ only due to the difference in the amount of structural distortion~\cite{longo1968magnetic}. It has been proposed~\cite{PhysRevB.77.214410} and subsequently experimentally demonstrated that modifying the Ru-O-Ru bond angle and orientation by either chemical substitution ~\cite{he2001disorder, maignan2006ferromagnetism} or tensile strain~\cite{tripathi2014ferromagnetic} can induce ferromagnetism in this material. The exact magnetic nature of \CRO films is debated~\cite{PhysRevB.54.8996,longo1968magnetic} - it has been suggested that the conflicting observations by different groups can be, at least to some degree, be explained by differences in the samples used for the studies~\cite{PhysRevB.77.214410}. Our measurements show conclusively that the ferromagnetic order seen in these materials depends critically on the level of tensile strain present in these films.

	\begin{figure}[tbh]
		\begin{center}
			\includegraphics[width=0.48\textwidth]{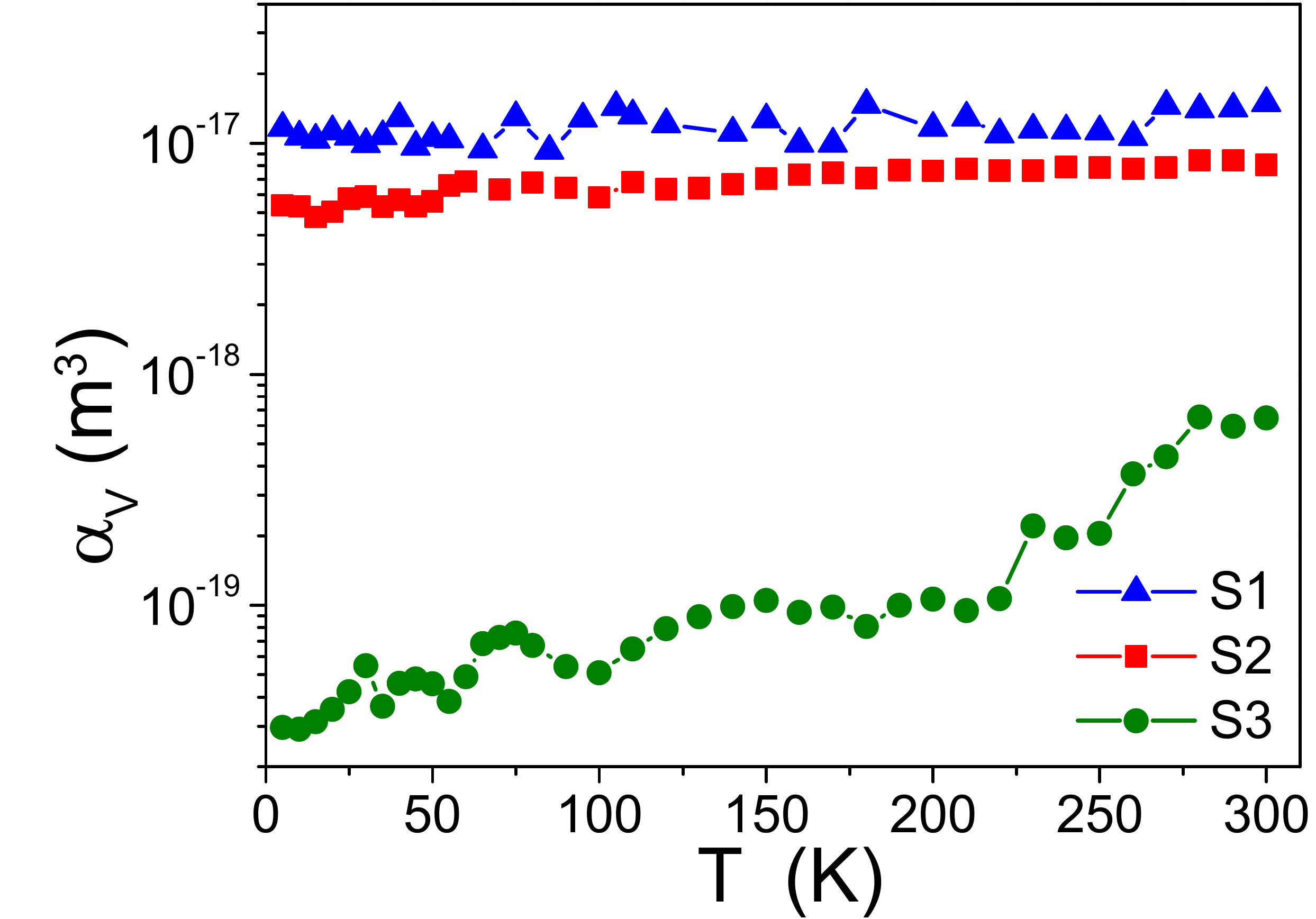}
			\small{\caption{ Linear-log plot of $\alpha_V$ as a function of temperature for the three \CRO films. Note that the noise scales with the amount of disorder (quantified by the RRR) in the films. \label{fig:Hooge}}}
		\end{center}
	\end{figure}
	
	To understand further the effect of film microstructure on its transport properties we have measured low frequency resistance fluctuations (noise) in all three films using a digital signal processing (DSP) based ac 4-probe technique~\cite{scofield1987ac} (details of the measurement as well as data analysis techniques can be found in our previous publications~\cite{PhysRevLett.111.197001, PhysRevB.67.174415, PhysRevB.90.115153}). This technique allows simultaneous measurement of the background noise as well as the noise from the  sample under study. The power spectral density (PSD) $S_R(f)$ of resistance fluctuations was evaluated from the measured time series of resistance fluctuations accumulated using a fast analog to digital conversion card (ADC). The PSD was seen to have a power law dependence on the spectral frequency $f$, $S_R(f) \propto f^{-\alpha}$ with the value of $\alpha$ very close to unity in all cases.  In figure~\ref{fig:Dutta} we have plotted the values of $\alpha$ = -($\partial$ ln$S_R$(f)/$\partial$~lnf) obtained from the slope of the PSD $S_R(f)$. In all three films $\alpha$ is seen to be almost independent of temperature down to 4~K. 
	
	\begin{figure}[tbh]
		\begin{center}
			\includegraphics[width=0.48\textwidth]{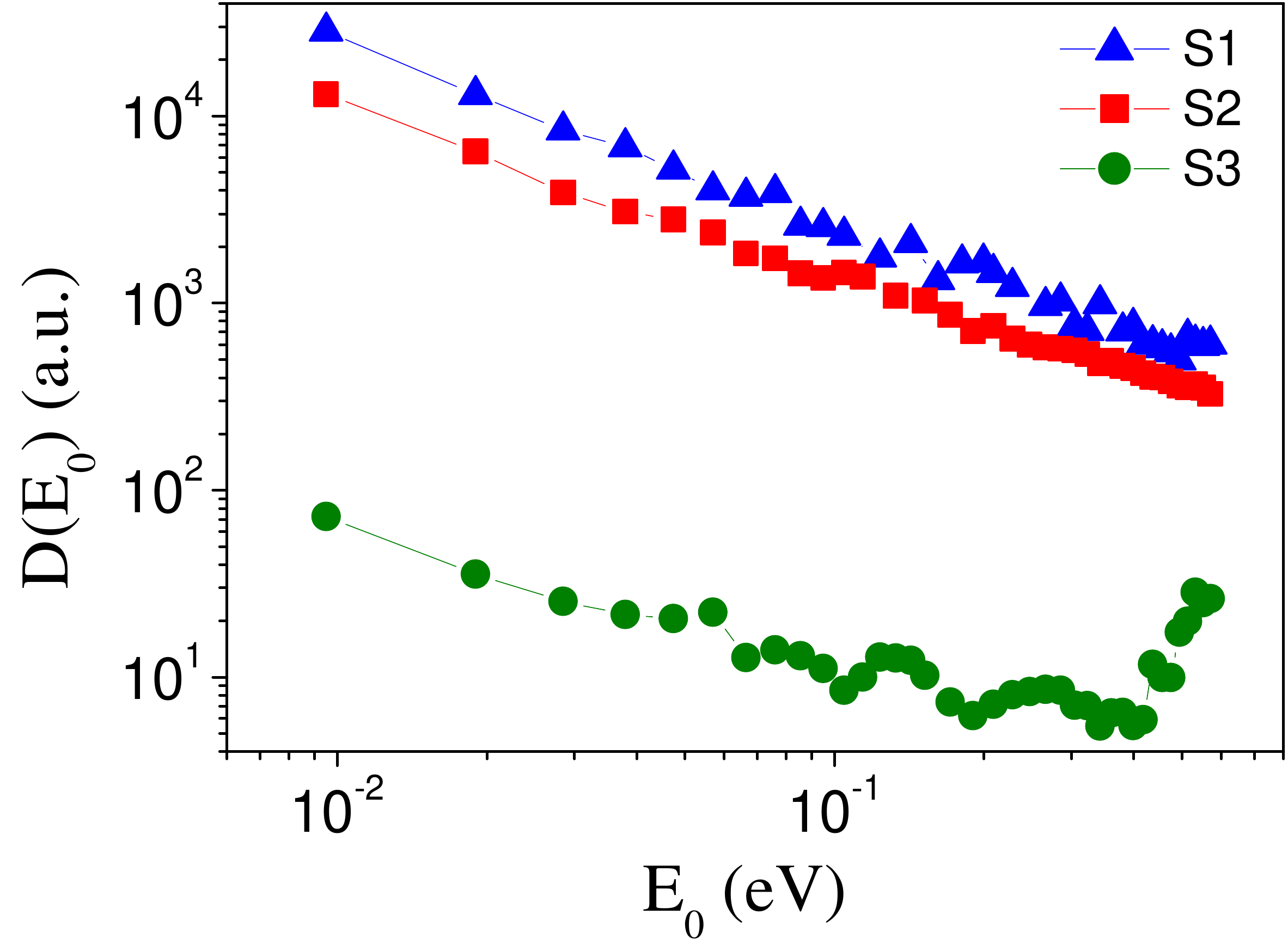}
			\small{\caption{ Distribution of activation energies $D(E_0)$ of thermally activated processes calculated from the temperature dependence of the noise, plotted as a function of the acitvation energies for these processes.   \label{fig:Dutta1}}}
		\end{center}
	\end{figure}
	
	\begin{figure}[tbh]
		\begin{center}
			\includegraphics[width=0.32\textwidth]{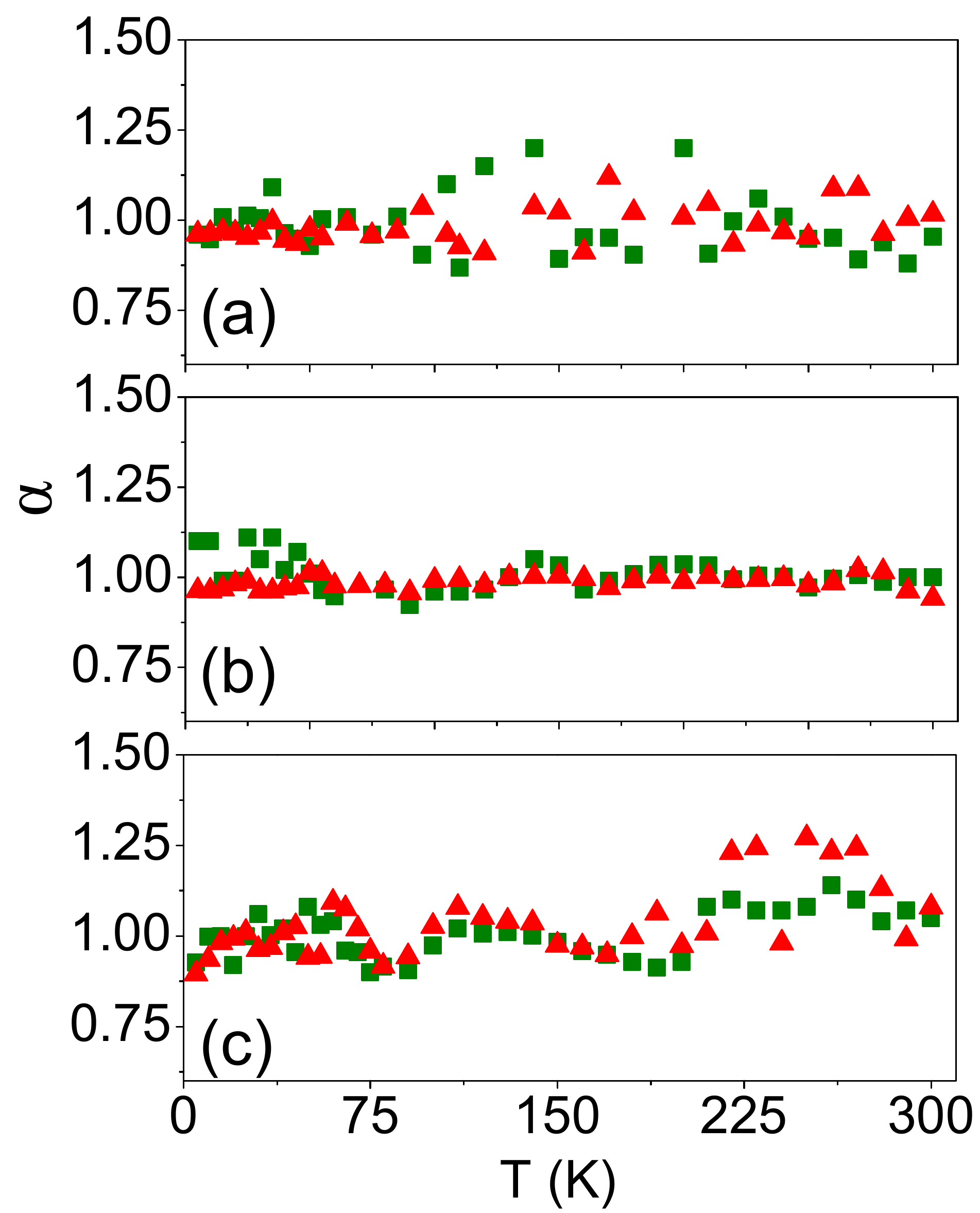}
			\small{\caption{ Plots of the exponent $\alpha$ for the three films (a) S1, (b) S2 and (c) S3. Red filled traingles are the values extracted from the temperature dependence of the relative variance of resistance fluctuations using the Dutta-Horn model. The olive filled squares are the values extracted from frequency dependence of the power spectral density of reistance fluctuations, $\alpha$ = -($\partial$ ln$S_R$(f)/$\partial$~ln$f$). \label{fig:Dutta}}}
		\end{center}
	\end{figure}

	To compare the magnitude of noise seen in the three different films  we used the frequency independent parameter $\alpha_V$ defined as:
	\begin{equation}
	\alpha_V=V\frac{fS_R(f)}{R^2}
	\end{equation} 
	where $V$ is the volume of the sample. In figure \ref{fig:Hooge} we have plotted $\alpha_V$ as a function of temperature for the three different \CRO films. For metallic single component continuous films, resistance noise arises primarily from local fluctuations in the sheet resistance~\cite{scofield19851} and can be associated with the scattering of charge carriers by extrinsic defects or impurities. The noise in this case is known to scale as the residual resistivity ratio (RRR) of the film~\cite{scofield19851}. This is consistent with our observations that near room temperatures the noise in S1 is almost an order of magnitude larger than that of S3 with the noise magnitude of S2 lying in between these two values. The temperature dependence of noise in such systems can be attributed to thermally activated processes with a distribution of relaxation times $\tau_o$~\cite{PhysRevLett.43.646}. The distribution $D(E_0)$ of the activation energies $E_0$ leading to these processes is related to the power spectral density of resistance fluctuations  as~\cite{PhysRevLett.43.646,dutta1981low, PhysRevB.31.1157} as:
	\begin{equation}
	D(E_0) \sim \frac{\omega S_V(\omega,T)}{k_BT}
	\label{eqn:DH1}
	\end{equation}  
	where $\omega=2\pi f$, $E_0 = k_BTln(2\pi f \tau_0)$ and $\tau_0$ is the attempt frequency of the activated process. Typical values of $\tau_0$ are of the order of $10^{-14}$ sec (of the order of an inverse phonon frequency)~\cite{PhysRevLett.43.646}. The values of $D(E_0)$   calculated using eqn.~\ref{eqn:DH1} are plotted in figure~\ref{fig:Dutta1}.  The disorder-free nature of S3 is reflected in the much lower defect density levels present in it as compared to S1 or S2.

	If the resistance fluctuations indeed arise due to the thermally activated  fluctuations of local defects then the value of $\alpha$ can be extracted independently from the temperature dependence of the noise from the Dutta-Dimon-Horn model~\cite{dutta1981low} using the following relation:
	\begin{equation}
	\alpha(\omega,T)=1-\frac{1}{ln(\omega \tau_0)}(\frac{\partial lnS_R(\omega,T)}{\partial lnT}-1)
	\label{eqn:DH}
	\end{equation}
	In Figure \ref{fig:Dutta} we have plotted the values of $\alpha$ extracted from this model as a function of temperature for all three films. It can be seen that these values of $\alpha$ match  very well with the values calculated from the frequency dependence of the power spectral density of resistance fluctuations, $\alpha$ = -($\partial$ ln$S_R$(f)/$\partial$~lnf). This lends credence to the idea that the dominant source of resistance fluctuations in these materials are local scatterers.
	
	\section{Conclusions}
	
	To conclude, we have carried out extensive structural and transport studies on three films of \CRO grown under different conditions. We find that when all other growth parameters are kept fixed, oxygen partial pressure affects very strongly the preferred orientation and surface roughness of the films.  This in turn has profound implications on the magnetic ordering in this material - in the films having large disorder we observe signatures of ferromagnetic behaviour from characteristic hysteresis loops in the MR curves. For the  best quality films on the other hand, the MR does not show any hysteresis. The electrical transport properties is strongly affected by the  oxygen  growth pressure as seen from the values of resistivity and the resistance fluctuations.  We could verify that the resistance fluctuations in these materials arise due to thermally activated fluctuations of local defects.  
	
	\section*{Acknowledgements}  
	
	We acknowledge funding from Nanomission, Department of Science \& Technology (DST) and HRDG, CSIR, Govt. of India.

%	\section*{References}
	
	\bibliographystyle{elsarticle-num}
	%\bibliography{cro_noise}

%\bibliographystyle{apsrev}
%\bibliographystyle{apsprl}
%\bibliographystyle{apsrev4-1}

\end{document}